\documentclass{emulateapj}
\begin{document}
\title{The Submillimeter Bump in Sgr A* from Relativistic MHD Simulations}
\shorttitle{Modeling Sgr A* mm Emission}
\shortauthors{Dexter et al}
\author{Jason Dexter}
\affil{Department of Physics, University of Washington, Seattle, WA 98195-1560, USA}
\email{jdexter@u.washington.edu}
\author{Eric Agol}
\affil{Department of Astronomy, University of Washington, Box 351580, Seattle, WA 98195, USA}
\author{P. Chris Fragile}
\affil{Department of Physics \& Astronomy, College of Charleston, Charleston, SC 29424, USA}
\author{Jonathan C. McKinney}
\affil{Kavli Institute for Particle Astrophysics and Cosmology, Stanford University, Stanford, CA 94305-4060, USA}
\begin{abstract}
Recent high resolution observations of the Galactic center black hole allow for direct comparison with accretion disk simulations. We compare two-temperature synchrotron emission models from three dimensional, general relativistic magnetohydrodynamic simulations to millimeter observations of Sgr A*. Fits to very long baseline interferometry and spectral index measurements disfavor the monochromatic face-on black hole shadow models from our previous work. Inclination angles $\le 20^\circ$ are ruled out to $3\sigma$. We estimate the inclination and position angles of the black hole, as well as the electron temperature of the accretion flow and the accretion rate, to be $i={50^\circ}^{+35{}^\circ}_{-15{}^\circ}$, $\xi={-23^\circ}^{+97{}^\circ}_{-22{}^\circ}$, $T_e=(5.4 \pm 3.0) \times 10^{10}$K and $\dot{M}=5^{+15}_{-2}\times10^{-9} M_\odot \mathrm{yr}^{-1}$ respectively, with 90\% confidence. The black hole shadow is unobscured in all best fit models, and may be detected by observations on baselines between Chile and California, Arizona or Mexico at $1.3$mm or $.87$mm either through direct sampling of the visibility amplitude or using closure phase information. Millimeter flaring behavior consistent with the observations is present in all viable models, and is caused by magnetic turbulence in the inner radii of the accretion flow. The variability at optically thin frequencies is strongly correlated with that in the accretion rate. The simulations provide a universal picture of the $1.3$mm emission region as a small region near the midplane in the inner radii of the accretion flow, which is roughly isothermal and has $\nu/\nu_c \sim 1-20$, where $\nu_c$ is the critical frequency for thermal synchrotron emission.
\end{abstract}
\maketitle

\section{Introduction}

Due to its proximity, the compact radio source at the Galactic center (Sgr A*; \citealt{balick1974}) is the most intensively studied supermassive black hole candidate. Using  very long baseline interferometry (VLBI) at 1.3mm, \citet{doeleman2008} detected source structure on event-horizon scales. Future measurements may detect the black hole shadow, providing the first direct evidence of an event horizon \citep{bardeen1973,falcke}.

Sgr A* is moderately variable in the radio and millimeter \citep{zhao2003,eckart2008sim,marrone2008,li2009,yusefzadeh2009} with order of magnitude flares in the IR and X-ray \citep{baganoff2001,genzel2003,ghez2004}. Its spectral energy distribution (SED) rises from the radio to a millimeter peak (``submillimeter bump"). \citet{baganoff2003} detected a much lower quiescent luminosity in the X-ray, while a quiescent infrared state has not been definitively detected. The submillimeter bump is thought to arise from relativistic, thermal electrons in the innermost portion of a hot, thick, advection-dominated accretion flow (ADAF; \citealt{narayanyi1995,yuanquataert2003}). The radio spectrum can be explained either by a small amount of the internal energy being injected into nonthermal electrons in the accretion flow \citep{yuanquataert2003} or by a short, mildly relativistic, optically thick jet \citep{falckemarkoff2000}. The quiescent X-ray emission is dominated by bremsstrahlung from cooler electrons further from the black hole, near the Bondi radius.

\citet{broderickloeb2006} modeled the infrared flaring as being due to orbiting inhomogeneities (``hot spots") in the inner radii of the accretion flow. Recently, correlated multiwavelength flares have instead favored an adiabatically expanding blob model \citep{yusefzadeh2009}, which can explain the observed time lags between the infrared, X-ray and millimeter flares. The jet emission model can explain the time lags between flaring events in the radio and millimeter \citep{falckeetal2009}.

The ADAF, jet, hotspot and expanding blob models for Sgr A* emission provide a variety of competing pictures of the accretion flow. These models are either completely non-relativistic or modified later to be consistent with relativity. They neglect the magnetic fields responsible for outward angular momentum transport and accretion via the magnetorotational instability (MRI, \citealt{velikhov1959,chandrasekhar1960,mri}). None can self-consistently account for quiescent and flaring behavior.
 
Magnetohydrodynamic (MHD) simulations can provide a more physical description of the accretion flow. They have been used to model the synchrotron emission from Sgr A*, either in three spatial dimensions with a pseudo-Newtonian potential \citep{ohsuga2005,goldston2005,huang2009,chan2009} or in two dimensions in full general relativity \citep{noble2007,moscibrodzka2009,hilburn2009}. Non-relativistic simulations are especially inappropriate for modeling the millimeter emission, which originates in the innermost portion of the accretion flow where relativistic effects are strongest. Axisymmetric simulations cannot sustain the MRI, and cannot accurately model variability.

The submillimeter bump is of particular interest. This is the frequency range where VLBI can both resolve the black hole event horizon, and where the measurements are no longer dominated by interstellar scattering. Unlike the radio, IR, and X-ray emission the dominant emission mechanism and electron distribution function is known in the millimeter. \citet{broderick2009} fit a non-relativistic, semi-analytic radiatively inefficient accretion flow (RIAF) model to spectral and mm VLBI data. \citet{dexter2009} fit time-dependent images of millimeter synchrotron emission from a three-dimensional GRMHD code to the VLBI data from \citet{doeleman2008}. In this work, we expand that analysis in several respects. We produce images from additional three-dimensional GRMHD simulations, including total energy conserving simulations from \citet{mckinneyblandford2009} and simulations at low black hole spins from \citet{fragileetal2009}. Images are fit to the spectral index measurements from \citet{marronephd} as well as the VLBI data from \citet{doeleman2008}, and two-temperature models of the accretion flow are considered.

\begin{table*}
\caption{\label{simtable}Key to the Simulations}
\begin{scriptsize}
\begin{center}
\begin{tabular}{lccccc}
        \tableline
	\tableline
 Name &  Spin (M) & Energy Scheme & $\Delta t$ (Used in Modeling) & Initial Conditions & Reference\\
        \tableline
0h & 0.00 & Internal & 7900M (4600M) & \citet{devilliers2003} & \citet{fragileetal2009} \\
50h & 0.50 & Internal & 7900M (4600M) & \citet{devilliers2003} & \citet{fragileetal2009} \\
90h & 0.90 & Internal & 7900M (4600M) & \citet{devilliers2003} & \citet{fragile2007} \\
MBD & 0.92 & Total & 3500M (2000M) & \citet{gammie2003} & \citet{mckinneyblandford2009} \\
MBQ & 0.94 & Total & 5500M (3000M) & \citet{gammie2003} with quadrupolar field & \citet{mckinneyblandford2009} \\
	\tableline
\end{tabular}
\end{center}
\end{scriptsize}
\end{table*}

\section{Simulations}
\label{sims}
Relativistic, three dimensional simulations are necessary to produce a physical, realistic description of the time-dependent accretion flow structure. Axisymmetric simulations cannot sustain turbulence and  exaggerate variability. Previous studies using 2D simulations have time-averaged the fluid structure, but even then the statistical steady state is not correct due to the decay of the MRI on the local orbital time. The ideal approach would be to use a single code to run a grid of 3D simulations over black hole spin, initial magnetic field configuration and initial torus location and geometry. However, given the formidable computational expense of such a project, we instead leverage a subset of existing 3D GRMHD simulations from two different groups. The simulations and their parameters are listed in Table \ref{simtable}.

The two sets of simulations use different initial conditions and energy evolution equations. Simulations from \citet{fragile2007,fragileetal2009} evolve internal energy and use the initial torus configuration from \citet{devilliers2003}, whereas those from \citet{mckinneyblandford2009} conserve total energy and use an initial torus placed much closer to the black hole \citep{gammie2003}. The two simulations from \citet{mckinneyblandford2009} differ slightly in black hole spin, as well as in initial magnetic field configuration. The simulation labeled MBD uses a standard, small-scale dipolar loop. The MBQ simulation has a large-scale quadrupolar field, which leads to a stronger MRI, slightly higher magnetic pressures and a larger accretion rate.

Total energy conserving simulations are probably more appropriate for modeling Sgr A*, since its inferred radiative efficiency is relatively low ($\epsilon \sim .001-.01$). \citet{dexter2009} computed a radiative efficiency from numerical energy losses of $\epsilon \sim .1$ for the 90h simulation, and a bolometric ``luminosity" exceeding that of Sgr A* by a factor of a few. However, few energy conserving three dimensional GRMHD simulations have been run to date \citep{shafee2008,noble2009,noblekrolik2009,mckinneyblandford2009,noblekrolik2010}. Only \citet{mckinneyblandford2009} simulated a thick, advection-dominated accretion flow. The simulations used here do not include radiative cooling, which is likely a good approximation for an ADAF.. Non-conservative simulations, where a significant amount of energy is lost to artificial numerical cooling, are still advection-dominated.

For comparison, we have also run a set of axisymmetric simulations using the publicly available HARM code \citep{gammie2003,noble2006} at black hole spins $a/M=0.00$, $0.50$, $0.75$, $0.90$, $0.92$, and $0.94$ ($G=c=1$ is used throughout this paper). They are not used for fitting to observations due to the problems with axisymmetric simulations described above. Instead, they give a sense of the spin dependence of total energy conserving simulations, which helps to break the degeneracies between black hole spin and initial conditions in our 3D simulations. They also provide an opportunity for direct comparison with results from previous work using similar simulations.

\section{Modeling}
\label{modeling}
Relativistic synchrotron radiation is computed from the simulations by tracing rays backwards from an observer's camera at infinity through the accretion flow. Each ray constitutes a pixel of the image, which are spaced uniformly on a rectangular grid. The points along each ray are assumed to be null geodesics of the Kerr spacetime, and their trajectories are computed using the public code \texttt{geokerr} \citep{dexteragol2009}. The code computes all coordinates of the geodesics, and the ray tracing procedure is fully time-dependent. Tracing rays backwards is ideal for Sgr A*, where the images themselves are observables.

We find that an image resolution of $150$x$150$ pixels (rays), with maximum spacing of $.2$M in impact parameter at infinity is sufficient for total flux  convergence to $\simeq 1\%$ relative to $300$x$300$ or $600$x$600$. The largest single pixel errors are $\simeq 5-10\%$, concentrated near the circular photon orbit. We find $\sim 400$ points on each geodesic to be adequate for all models. These points are spaced evenly in $1/r$, except near radial turning points, where \texttt{geokerr} uses the polar angle as the independent variable for better resolution.

The unpolarized synchrotron emission coefficient is computed from Leung et al (2010; submitted) by linearly interpolating simulation fluid variables to the points along each ray. For reasons discussed below, we often use data from the nearest time step rather than interpolating in the time direction. The absorption coefficient is obtained from Kirchoff's Law. The gas pressure, magnetic field and particle density are taken from the simulations and scaled to physical units with a single free parameter representing the total torus mass. This parameter fixes the mass accretion rate. Only emission from fluid within $r=25$M is used for computing images, and only the last $\sim 60\%$ of the simulation data is used, after transients from the onset of turbulence have died down (see Table \ref{sims}). The millimeter emission in the models considered is always concentrated well within both our radial boundary and the initial pressure maximum for all simulations.

The simulations evolve the total gas pressure of electrons and ions, while only the electron temperature is used in calculating the emergent radiation. We follow \citet{goldston2005,moscibrodzka2009} and generate images using two-temperature models with constant ratios of $T_i/T_e$, which is then another free parameter. 

The models are fit to both the VLBI observations at $1.3$mm from \citet{doeleman2008} and the spectral index measurements between $.4$mm and $1.3$mm from \citet{marronephd}. This gives the probability of observing the measured values from a particular model. Following \citet{broderick2009}, this distribution is converted to the probability distribution as a function of our parameters given the observations using Bayes' theorem. The result is two separate probability distributions: $p_v(\dot{M}_v,t_v,\xi,T_i/T_e,a)$ and $p_s(\dot{M}_s,t_s,
T_i/T_e,a)$, where $t$ is the observer time, $\dot{M}$ is the time-averaged accretion rate, $\xi$ is the orientation of the black hole spin axis projected on the sky, $T_i/T_e$ is the ion electron temperature ratio, and $a$ is the spin value of the simulation used. The subscripts $v$ and $s$ refer to the VLBI and spectral fits respectively, and the simulations differ in more respects than just the value of the black hole spin. To produce parameter estimates, we average over observer time and marginalize over accretion rate and multiply the resulting distributions. To estimate the value of a single parameter, we then marginalize the combined probability distributions over the other parameters. To estimate the accretion rate, we average over observer time, combine the resulting distributions, and then marginalize over all other parameters.

For each simulation, images are produced over a grid of the above parameters at $.4$mm and $1.3$mm. VLBI fits are done as in \citet{dexter2009} following \citet{broderick2009}. Visibility amplitudes are calculated by taking the absolute value of the Fourier transform of the $1.3$mm image averaged over 10 minute intervals to match the observations. The visibility is then rotated to the desired sky orientation, and multiplied by an elliptical Gaussian to account for interstellar scattering as done in \citet{fish2009} using the fits from \citet{bower2006}. The rotated, scatter-broadened visibility amplitude is then interpolated to the baseline locations of the VLBI measurements, and fit to their measured values.

The total fluxes at $.4$mm and $1.3$mm are time-averaged over $2.5$hr intervals to mimic the observations from \citet{marronephd}. We average the observed fluxes at $1.3$mm and spectral indices between $.4$mm and $1.3$mm over the four observational epochs, and fit the theoretical values to the averaged observational values. Uncertainties are determined by adding half the range of the individual measurements to the instrumental errors in quadrature. This procedure gives a $1.3$mm flux of $3.75 \pm .48$ Jy and a spectral index of $-.18 \pm .34$, where the uncertainties are one sigma. Note that the $1.3$mm flux during these observations was $\sim 50\%$ higher than during the VLBI measurements from \citet{doeleman2008}. Image intensities are converted to fluxes using a black hole mass of $4 \times 10^6 M_\odot$ and a distance of $8$ kpc.

Since the individual measurement errors from \citet{marronephd} are smaller than the flux differences between the measurements, the data provide information about source variability as well as a time-averaged value. Instead of using an averaged value as described above, we could instead fit to each measurement separately. Our light curves have been analyzed both ways. Using this second method provides slightly tighter constraints for individual models, but the parameter estimates and best fit models discussed below are identical between the two methods.

\begin{figure}
\epsscale{1.1}
\plotone{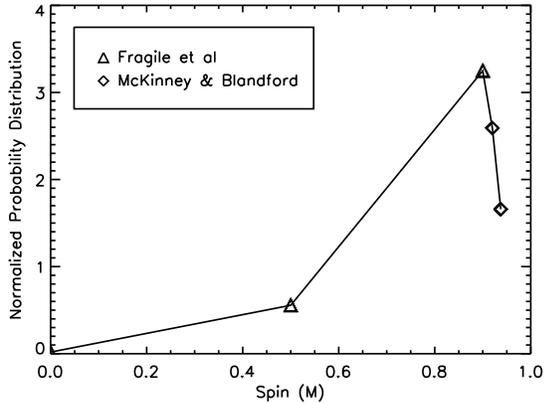}
\caption{\label{pva}Normalized probability density as a function of black hole spin, marginalized over inclination angle and $T_i/T_e$. Triangles denote the first three entries in Table \ref{simtable}, from \citet{fragile2007,fragileetal2009}. Diamonds represent the last two entries in Table \ref{simtable} from \citet{mckinneyblandford2009}.}
\end{figure}

\begin{figure}
\epsscale{1.1}
\plotone{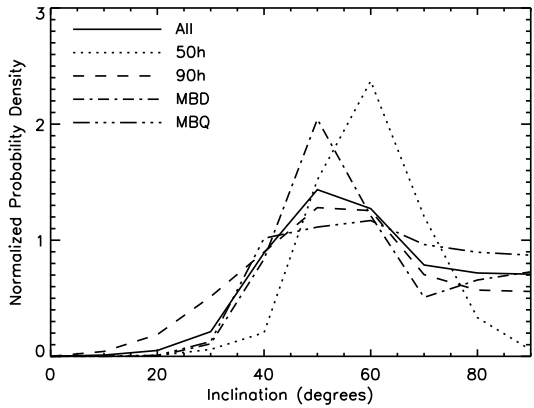}
\caption{\label{pvi}Normalized probability density as a function of observer inclination, for each simulation separately and marginalized over simulation (black hole spin). All curves are marginalized over $T_i/T_e$. The 0h simulation isn't shown due to its negligible contribution to the overall curve, but is similar to that of 50h.}
\end{figure}

\begin{figure}
\epsscale{1.1}
\plotone{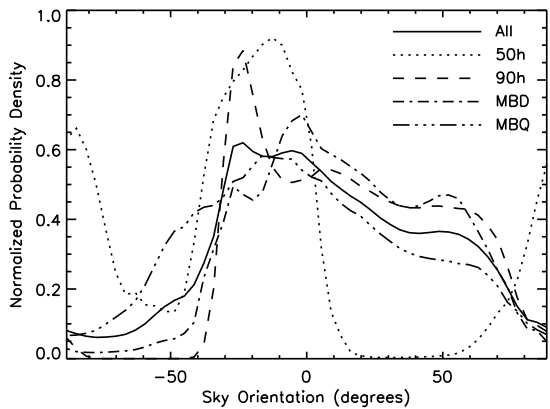}
\caption{\label{pvx}Normalized probability density as a function of sky orientation, for all simulations (solid) as well as each simulation individually.}
\end{figure}

\begin{figure}
\epsscale{1.1}
\plotone{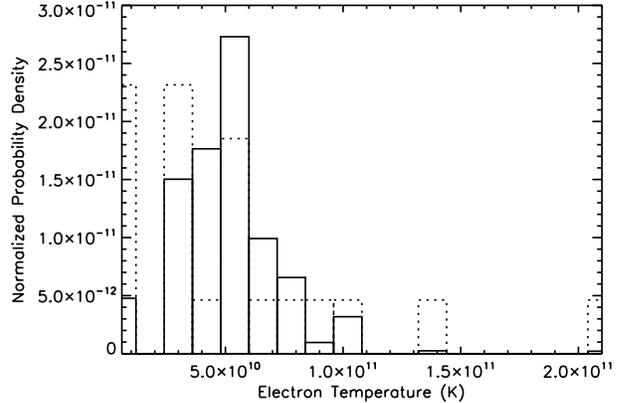}
\caption{\label{electron}Normalized probability density as a function of electron temperature, marginalized over inclination angle and simulation (black hole spin). To create this plot, the grid of simulations over $T_i/T_e$ was binned by median electron temperature in the region of largest emissivity. Probability densities for models falling in each bin are added to make the histogram. The dotted histogram is the result of assigning equal probability to each model.}
\end{figure}

\begin{figure}
\epsscale{1.1}
\plotone{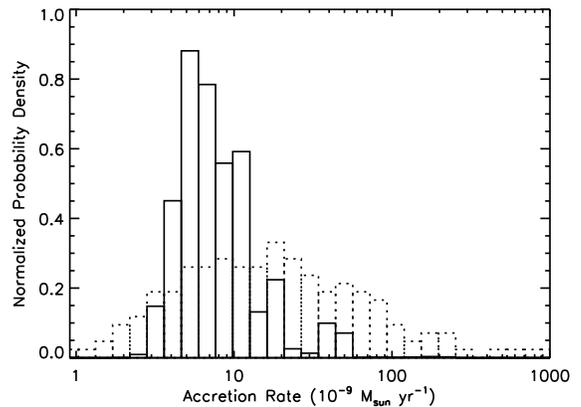}
\caption{\label{pvm}Normalized probability density as a function of accretion rate, marginalized over inclination angle and simulation (black hole spin). Probability densities for models with accretion rates falling in each bin are added to make the histogram. The dotted histogram is the result of assigning equal probability to each accretion rate sampled.}
\end{figure}

\section{Results}
\label{results}
The model of the millimeter emission from Sgr A* as realized in the simulations is of a compact, hot, magnetized accretion flow in the vicinity of a black hole. The millimeter is the peak of the spectrum and the frequency range which produces the majority of the radiation. It is produced by thermal electrons near the midplane of the disk in the inner radii, with typical values of $n \sim 5 \times 10^7 \mathrm{cm}^{-3}$, $B \sim 50$G, and $T_e \sim 5 \times 10^{10}$K. These typical values have been found by a number of modelers \citep{yuanquataert2003,goldston2005,moscibrodzka2009}.

As found in \citet{dexter2009}, the GRMHD simulations provide excellent fits to the observed millimeter emission from Sgr A* (reduced $\chi^2 \le 1.5$ frequently for all simulations). This is still true when we incorporate spectral index information as described above. However, the inclusion of that observational constraint and of additional simulations allow us to estimate some parameters of the accretion flow. Single parameter probability distributions are found by marginalizing the combined spectral and VLBI probability distributions over the others. To estimate confidence intervals, we use the technique described in \citet{broderick2009}. The probability distribution is integrated from its maximum value, $p_{max}$, down to a cutoff value $p_{min}$ so that the cumulative probability enclosed between $p_{min}$ and $p_{max}$ is equal to the desired confidence interval. Then the estimated value, lower and upper bounds of the parameter $x$ are $x(p_{max})$, $\mathrm{min}[x(p \ge p_{min})]$, and $\mathrm{max}[x(p \ge p_{min})]$ respectively.

\begin{figure*}
\epsscale{1.1}
\plotone{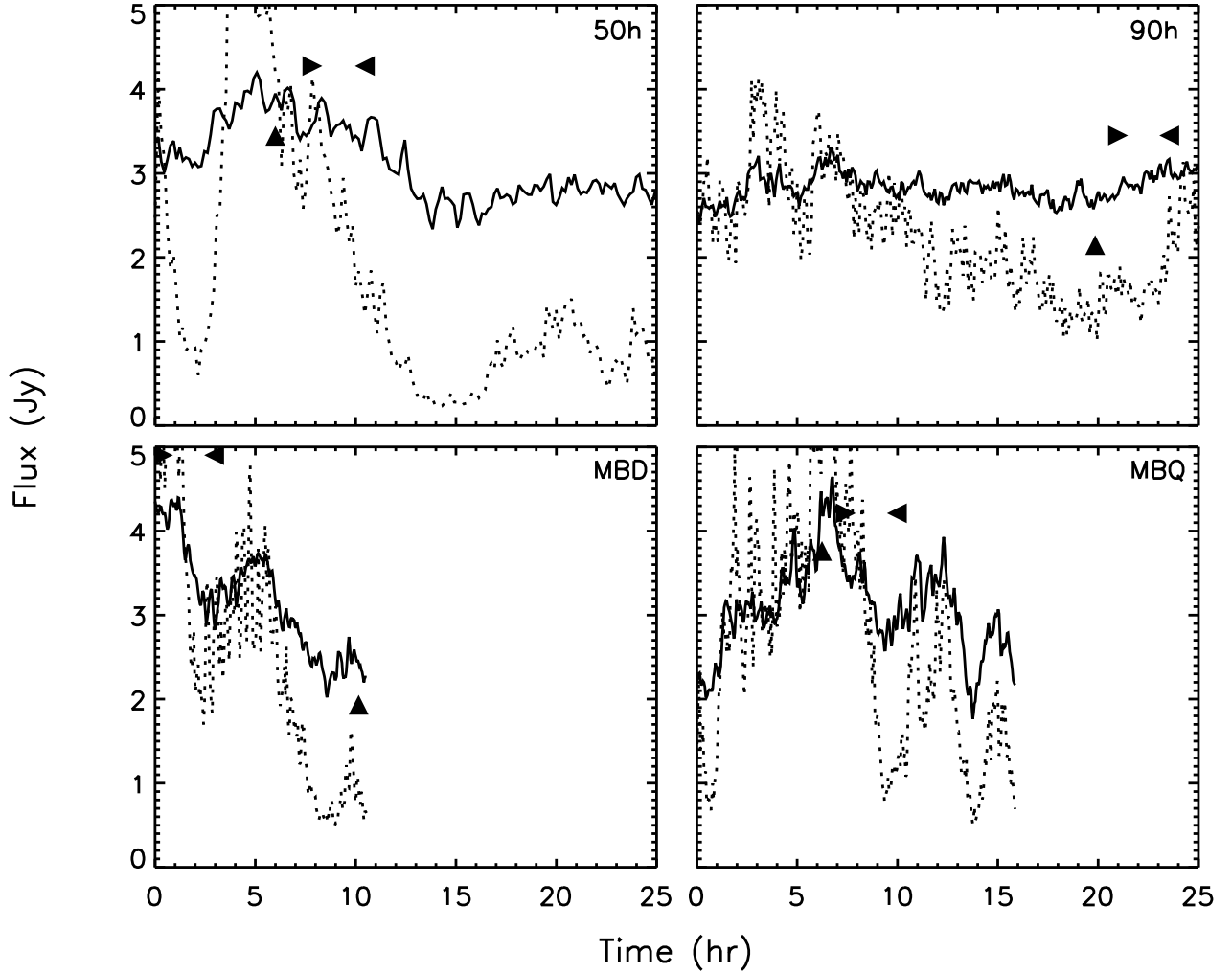}
\caption{\label{lcurves}Light curves for best fit parameters (see Table \ref{bestfit}) from each simulation at $1.3$mm (solid) and $.4$mm (dotted). The vertical arrow denotes the time of best fit to the VLBI measurements, while the horizontal arrows denote the averaged interval corresponding to the best fit to the spectral index observations. The accretion rate used for the plots is not necessarily the best fit, since that value can differ between the two constraints. The \citet{mckinneyblandford2009} simulations were run for shorter times since the torus was initially placed at smaller radius.}
\end{figure*}

\subsection{Parameter Constraints}

We estimate the electron temperature in the millimeter emission region, the accretion rate onto the black hole, the inclination angle of the accretion flow relative to the observer and the projection of the black hole spin axis on the sky. Unfortunately, our use of existing simulations precludes us from determining a constraint on the black hole spin. A plot of $p$($a$) is shown in Fig. \ref{pva}. The large peak at high spin is deceptive, and is due to the differences between the simulations rather than black hole spin. While it appears we might be able to rule out low spin simulations, the 0h simulation is probably a special case. Low electron temperatures in that simulation cause it to have a large photosphere ($r \sim 10$M) even when $T_i=T_e$, and therefore it fits the observations poorly. The axisymmetric HARM simulations, which are more similar to a 3D conservative simulation like MBD/MBQ, have larger electron temperatures at all spins and are optically thin for $T_i=T_e$. So although the 0h model is ruled out, low spin conservative simulations should have larger electron temperatures and provide better fits. We do not consider the 0h simulation for the remainder of the paper due to its negligible influence on the overall probability distributions.

All best fit models have similar probability distributions in inclination angle. These are shown in Fig. \ref{pvi}, along with the combined probability distribution after marginalizing over all other parameters. From this we estimate the inclination angle as $i={50^\circ}^{+35{}^\circ}_{-15{}^\circ}$ with $90\%$ confidence. This interval may change with more complete sampling of the spin and initial conditions parameter space; however the similarity of the curves from different simulations suggests that such changes will be minor. The face-on, black hole shadow fits at $i \lesssim 20^\circ$ found in \citet{dexter2009} are now ruled out to 3$\sigma$ confidence. One reason is that the spectral index constraint, which was not used in \citet{dexter2009}, disfavors low inclinations. Also, good fits to the VLBI data at low inclination occur infrequently in the MBD and MBQ simulations, which were not considered in the previous study. The VLBI measurements also depend on the sky orientation, $\xi$, the position of the black hole spin axis projected on the sky measured E of N. A plot of $p$($\xi$) is shown in Fig. \ref{pvx}. The distribution is quite broad, and we estimate a value of $\xi={-23^\circ}^{+97{}^\circ}_{-22{}^\circ}$ with $90\%$ confidence. Both the estimated values and the probability distributions shown here are in excellent agreement with \citet{broderick2009} (c.f. their Fig. 7).

For all simulations, the probability distribution over $T_i/T_e$ has a peak value between 1 and 10. The best fit $T_i/T_e$ rises with MHD temperature, so that roughly the same electron temperature fits best across all simulations. The probability distribution of emissivity averaged electron temperature is plotted in Fig. \ref{electron}. The dotted histogram shows the result of assigning equal probability density to each model. With better sampling, the distribution would likely be much more smooth. The electron temperature is estimated to be $(5.4 \pm 3.0) \times 10^{10}$K with $90\%$ confidence. As discussed below, our simulated light curves are consistent with an isothermal emission region. This electron temperature is consistent with electron temperatures found in previous work \citep{yuanquataert2003,goldston2005,sharma2007,moscibrodzka2009}.

Similarly, the accretion rate can be constrained from the joint fits despite the different ranges found for best fit parameters for each simulation (see Table \ref{bestfit}). The probability distribution over accretion rate is shown in Fig. \ref{pvm}, and we estimate its value to be $\dot{M}=5^{+15}_{-2}\times10^{-9} M_\odot \mathrm{yr}^{-1}$ with $90\%$ confidence. The probability distributions from the two observational constraints have been combined to produce the plot, despite the fact that the total flux is $\sim 50\%$ higher during the spectral index measurements. This procedure favors models with substantial variability. However, probability distributions from the two constraints separately are similar to the one in Fig. \ref{pvm}, with slightly broader ranges of allowed accretion rates and estimated values in agreement with that of the joint fit.

\begin{figure}
\epsscale{1.1}
\plotone{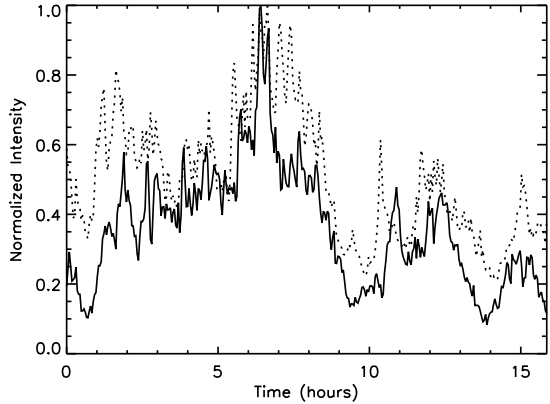}
\caption{\label{mdot}Light curve at $.4$mm (solid) and accretion rate at the inner boundary of the MBQ simulation (dotted) with best fit parameters (see Table \ref{bestfit}). Both quantities are scaled to their maximum value.}
\end{figure}

\begin{figure}
\epsscale{1.1}
\plotone{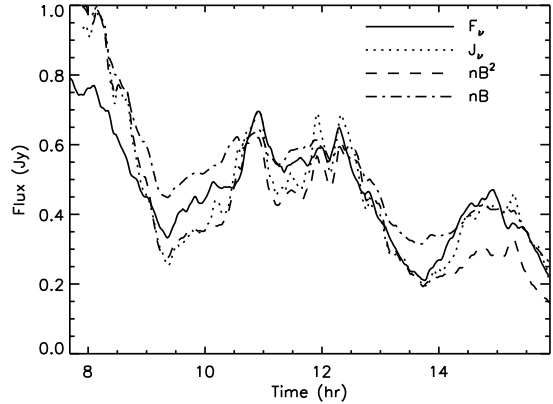}
\caption{\label{nb2}Last half of the $1.3$mm light curve of MBQ with $T_i/T_e=3$, $i=10^\circ$ ($F_\nu$) compared to the volume integrated synchrotron emissivity ($j_\nu$) and approximate volume integrated emissivities, $nB$ and $nB^2$ where $n$ is the electron density and $B$ is the magnetic field strength. The emissivities are normalized to their maximum values, while the light curve is scaled so that it matches the synchrotron emissivity at $\sim 11$ hours.}
\end{figure}

\begin{figure}
\epsscale{1.1}
\plotone{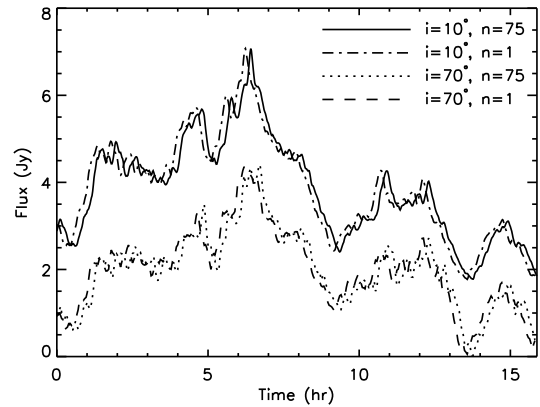}
\caption{\label{nload1}Light curves of MBQ at $i=10$ and $70$ degrees for $n=1,75$ where $n$ is the number of simulation time steps used to compute the intensity at a single observer time. $n=1$ neglects the effects of finite light travel time through the fluid.}
\end{figure}

\begin{figure}
\epsscale{1.1}
\plotone{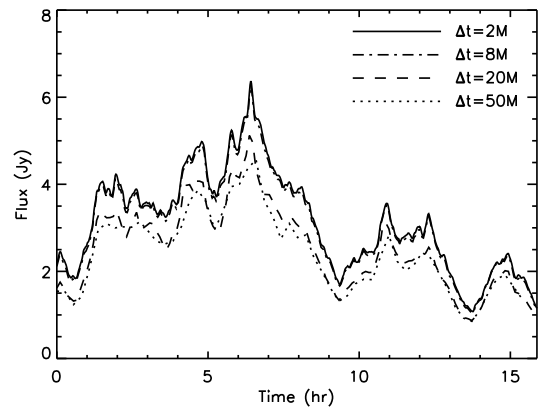}
\caption{\label{tdelays}Light curves of MBQ at $10$ degrees using simulation time steps separated by $\Delta t=2$, $8$, $20$ and $50$M, where $2$M is the value used for other MBD and MBQ light curves, while $\Delta t=4$M for the 50h and 90h simulations.}
\end{figure}

\subsection{Variability}

\citet{dexter2009} found that simulation 90h exhibited millimeter flares with $2.5$hr duration, $.5$hr rise times and $40\%-50\%$ amplitude, in agreement with observations of Sgr A*. Light curves at $1.3$mm (230 GHz) and $.4$mm (690 GHz) for best fit parameters from all viable simulations (excluding 0h) are shown in Fig. \ref{lcurves}. All light curves show millimeter flares that are consistent with the observations. The spectral index rises during the flares, often above zero, due to stronger variability at 690 GHz where the flow is completely optically thin. Only $1-2$ flares occur in each theoretical light curve, which is consistent with their observed frequency.

The flares are caused by increased magnetic field strength in the inner radii of the accretion flow where the synchrotron emissivity is highest. This is the case even in total energy conserving simulations, where the gas can be heated from reconnecting magnetic fields. Reconnection events, while present, do not seem related to the millimeter variability. Likewise, the variability is not produced by local inhomogeneities (hotspots), but rather by relatively global increases in the magnetic field strength in the inner radii. The light curves at $.4$mm are strongly correlated with variability in the accretion rate, as shown for simulation MBQ in Fig. \ref{mdot}. There are no robust lags found between the variability in the millimeter light curves and the accretion rate, and we cannot definitively determine the cause and effect. 

Since only the millimeter emission is modeled here, it is unclear if the flares produced are consistent with the observed x-ray, IR and millimeter flares \citep{eckart2008sim,marrone2008,doddseden2009,yusefzadeh2009}. At $1.3$mm, the flares are weaker at higher inclination, where they are attenuated by higher optical depth between the observer and the emission region. Due to the presence of a significant photosphere, the correlations with accretion rate variability are much weaker at longer wavelengths as well.

The $1.3$mm light curve shapes are closely reproduced by assuming the optically thin synchrotron emission comes from an isothermal region with a magnetic field satisfying $\nu/\nu_c \sim 1-20$, where $\nu_c=6.27\times10^{18} B (kT)^2$ (cgs) is the critical frequency for synchrotron emission. Then the emissivity is $j_\nu \sim nB^\alpha$, with $1 \le \alpha \le 2$. The light curve from simulation MBD is compared to the integrated synchrotron emissivity as well as these simplified isothermal models in Fig. \ref{nb2}. A low inclination light curve is used to lessen the impacts of Doppler beaming and optical depth. The flaring behavior is captured without any temperature fluctuations, and the variations are caused by those in the magnetic field strength and particle density. In models with significant photospheres, including most at high inclination, these simple formulas still work to describe the integrated synchrotron emissivity, but tend to be more variable than the light curves which include self-absorption.

As in \citet{dexter2009}, we compute radiative transfer in full general relativity, including accounting for time delays along geodesics. Including the time delays means that many different time steps of simulation data are used to compute the intensity along each geodesic. Fig. \ref{nload1} shows a comparison of light curves with and without accounting for time delays. The light curves are virtually identical at both low and high inclinations, except for a systematic shift in observer time, indicating the constant time lag between $r=25M$, where we begin following geodesics, to the emission region. 

This is surprising, especially given the results of \citet{noblekrolik2009}. They found significantly steeper power spectra at high frequencies when accounting for time delays. However, the simulation from that work used an artificial emissivity, lower time resolution ($\Delta t=20M$) and only a $\pi/2$ wedge in $\phi$. We've tried matching their $\pi/2$ wedge in $\phi$ and sampling with the same time resolution. In all cases, we find no systematic differences when accounting for light travel time through the accretion flow, and the errors due to neglecting time delays are at the $10\%$ level (Fig. \ref{nload1}), comparable to the overall uncertainty in the interpolation scheme or emissivity used. The largest errors occur at high inclination, where the light travel time to different parts of the emission region is largest. When using both the limited domain in $\phi$ and $\Delta t=20M$, the slope of the power spectrum is slightly steeper when using time delays. However the deviation is considerably smaller than that reported by \citet{noblekrolik2009}, possibly due to the differing emissivities or disk scale heights. They also included cooling self-consistently in the simulation.

Their use of a $\pi/2$ wedge in $\phi$ is predicated on the assumption that the dominant azimuthal spatial structure is in modes with $m\simeq4$ and higher \citep{schnittman2006}. We test this by expanding fluid variables in spherical harmonics for different simulations and time slices. In all cases, we find that the $m=1,2$ power is larger than at higher orders, as found previously by \citet{henisey2009} for a tilted simulation. This suggests that the use of restricted $\phi$ domains incorrectly approximates the spatial structure of a fully global simulation. In particular, we find that the power law index of the power spectrum is significantly steeper when using the full $2\pi$, indicating that using a limited wedge in $\phi$ introduces spurious variability on shorter timescales. Studying variability using simulations probably requires using the full $\phi$ domain.

We also caution that interpolating magnetic field vectors between time steps with insufficient time resolution results in systematic suppression of the total flux. This occurs when magnetic timescales are significantly shorter than the time step between simulation data dumps, causing interpolation between uncorrelated magnetic field vectors. For the MBD and MBQ simulations, $2-8$M time resolution is sufficient, as shown in Fig. \ref{tdelays}. However, when interpolating magnetic fields between time steps in the 90h simulation, the flux is systematically $\sim10\%$ lower with a time spacing of 4M. Our radiative transfer code uses time steps equal to an integer multiple of the simulation time step so that these errors cannot cause spurious features in the light curves. Regardless, we use data from the nearest time step rather than interpolating magnetic fields when the time steps are too large.

\begin{table}
\caption{\label{bestfit}Best Fit Model Parameters}
\begin{scriptsize}
\begin{center}
\begin{tabular}{lcccc}
        \tableline
	\tableline
 Name & Spin (M) & $\dot{M} (10^{-9} M_\odot \mathrm{yr}^{-1})$ & $i$ & $T_i/T_e$\\
        \tableline
50h & 0.50 & $20-40$ & $60^\circ$ & 1 \\
90h & 0.90 & $5-8$ & $50^\circ$ & 2\\
MBD & 0.92 & $4-10$ & $50^\circ$ & 3\\
MBQ & 0.94 & $6-14$ & $50^\circ$ & 6\\
	\tableline
\end{tabular}
\end{center}
\end{scriptsize}
\end{table}

\begin{figure*}
\epsscale{1.1}
\plotone{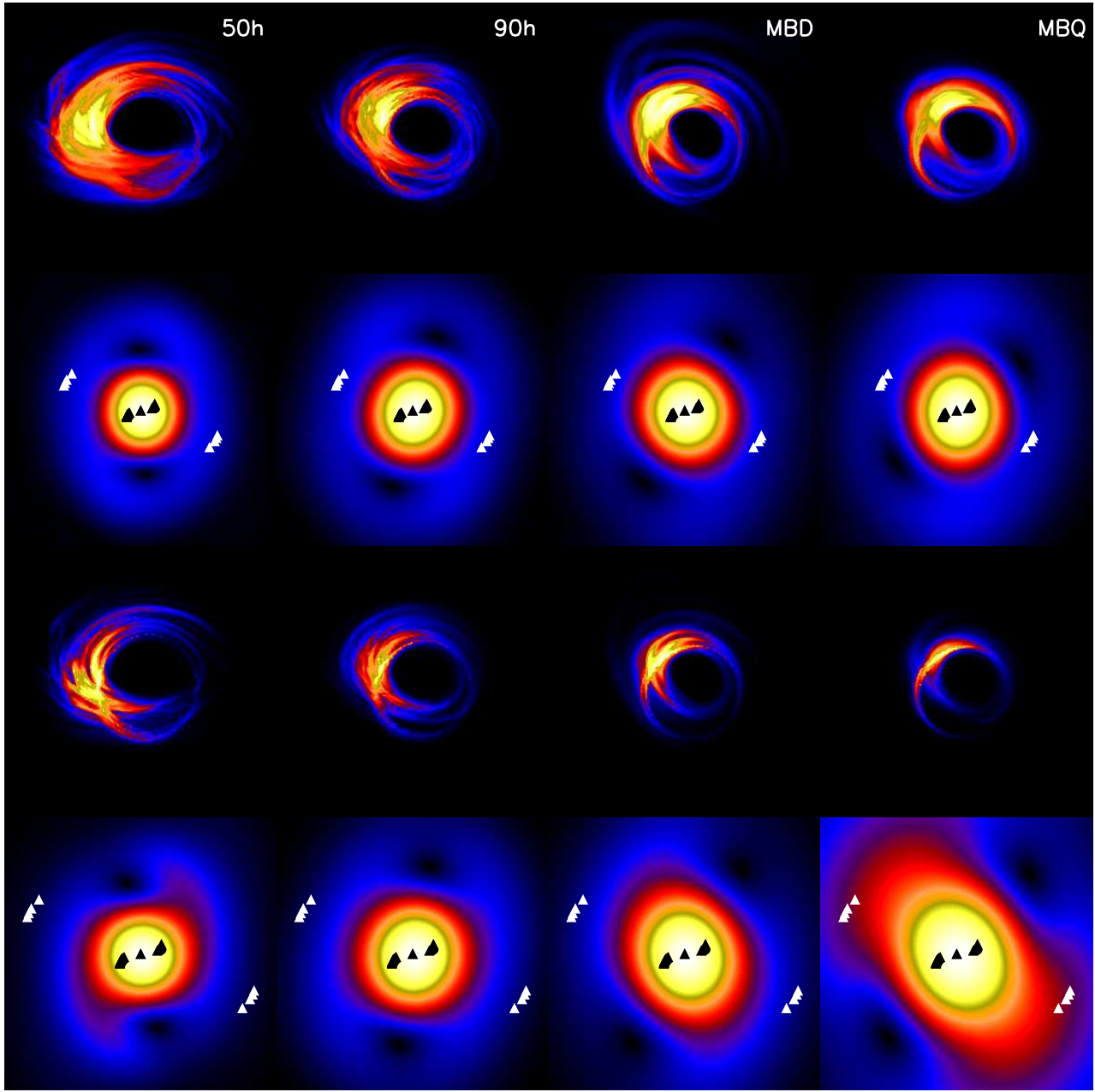}
\caption{\label{dintvis}Images and visibility amplitudes at $1.3$mm (first two rows) and $.87$mm (bottom two rows) for the best fit models. The first and third rows are images, while the second and fourth are the corresponding visibility amplitudes. All are rotated to their best fit positions. The uv-plane locations of the baselines used in \citet{doeleman2008} are over-plotted on the visibilities. The panel size is $150\mu$as, and 12 $G\lambda$ for the visibilities. The columns are labeled by simulation, and each panel is scaled to its maximum value. At $1.3$mm, the maximums are always $\simeq 2.4$Jy. However, the total flux can vary between simulations at $.87$mm. Since the images and visibilities form a Fourier transform pair, in general a larger image corresponds to a smaller visibility and vice versa.}
\end{figure*}

\begin{figure}
\epsscale{1.1}
\plotone{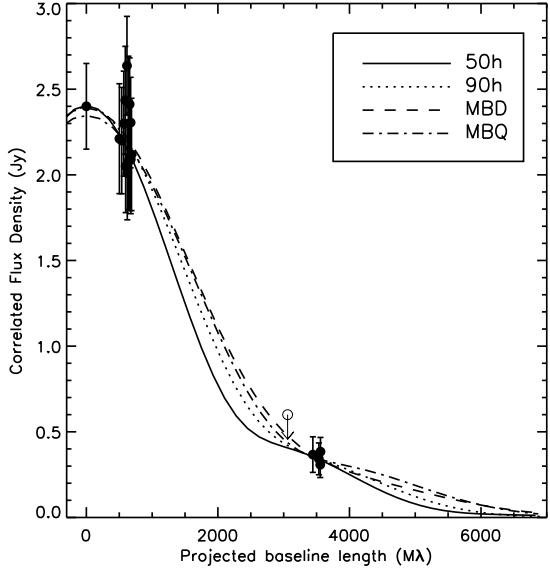}
\caption{\label{profiles}One dimensional visibility profiles along the line of the average long baseline location from \citet{doeleman2008}. Unlike the best fit models in \citet{dexter2009}, all of these profiles decrease monotonically with baseline length.}
\end{figure}

\begin{figure}
\epsscale{1.1}
\plotone{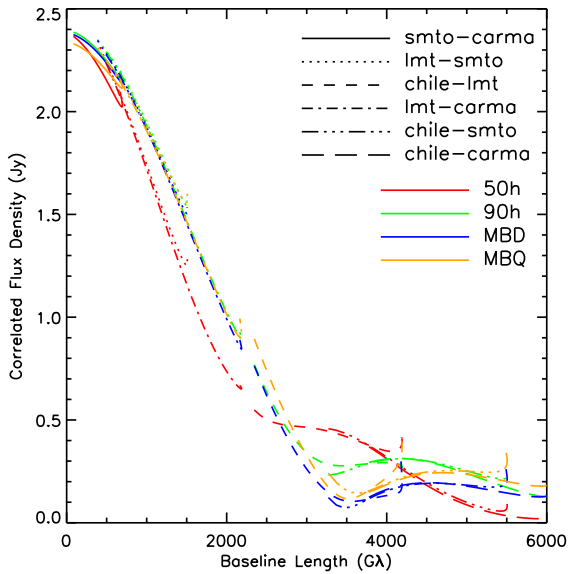}
\caption{\label{1dchile}Visibility amplitude as a function of baseline length for best fit models at $1.3$mm. The telescopes considered are in Arizona (SMTO), California (CARMA), Mexico (LMT) and Chile (APEX/ASTE/ALMA). The black hole shadow appears as a minimum in the 1D profile.}
\end{figure}

\begin{figure*}
\epsscale{1.1}
\plotone{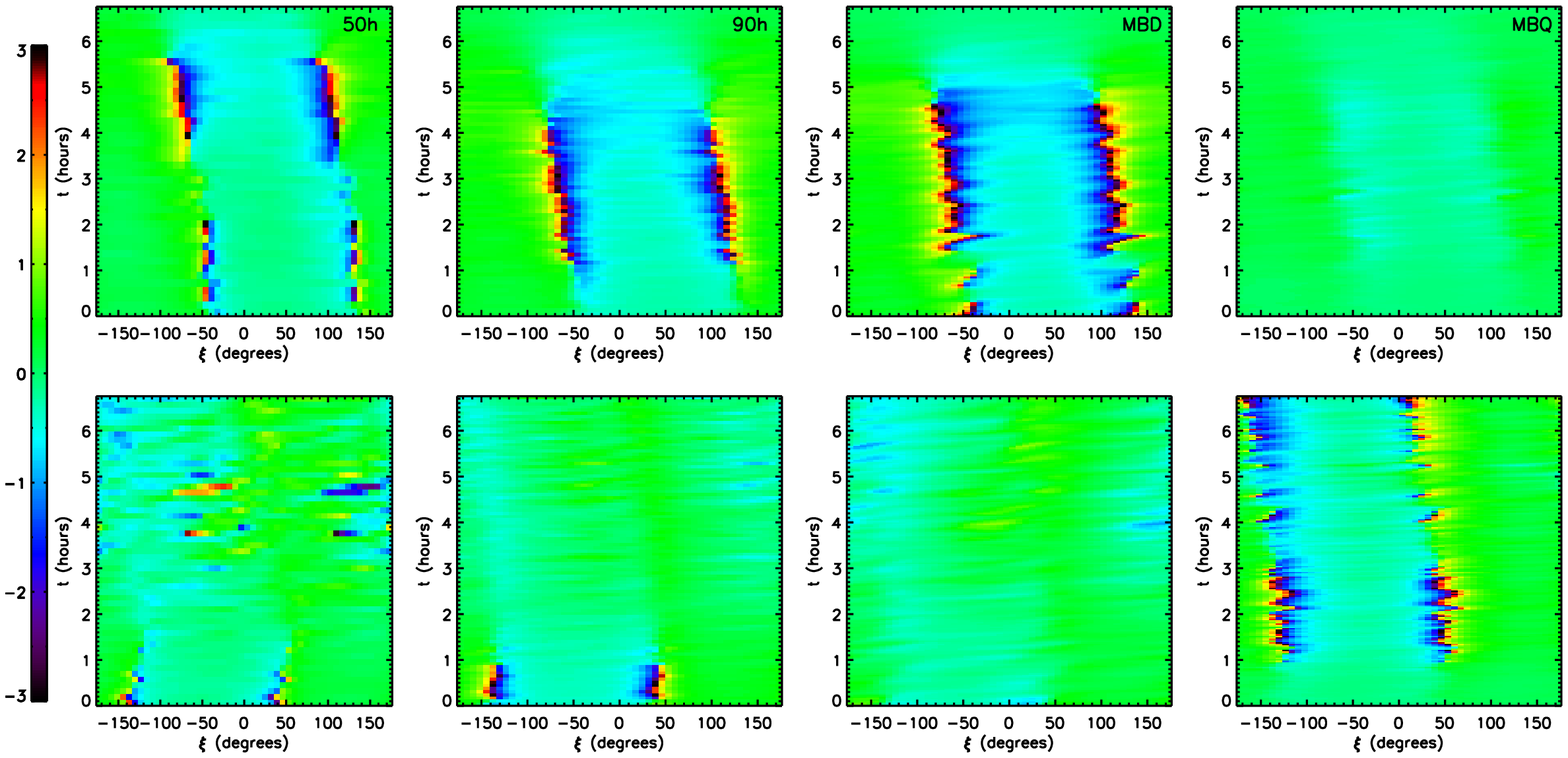}
\caption{\label{cphase}Closure phase as a function of observer time and sky orientation for best fit models on triangles of baselines between Arizona/California/Hawaii (top) and Arizona/Mexico/Chile (bottom).}
\end{figure*}

\subsection{Best Fit Models}

Table \ref{bestfit} lists the parameters for best fit models from the viable simulations (excluding 0h), as determined by the joint spectral index and VLBI fits. Images and visibilities from these models at $1.3$mm and $.87$mm are shown in Fig. \ref{dintvis}, at the time of best fit to the $1.3$mm VLBI observations. All best fit images are essentially the superposition of two crescents, due to the Doppler beaming of gas approaching the observer and the circular photon orbit. The shadow of the black hole is prominent along the vertical axis as a local minimum in the visibility amplitude. One dimensional visibility profiles, plotted along the line of the average location of the long baseline measurements from \citet{doeleman2008}, are shown in Fig. \ref{profiles}. In all models, the visibility profile decreases monotonically with baseline length along this axis. Therefore, the black hole shadow may not be seen by future measurements using the same telescopes as \citet{doeleman2008}.

The shadow is detectable from the two dimensional visibilities at baselines of similar length and roughly orthogonal orientation to those used by \citet{doeleman2008} in all best fit models. At $.87$mm, the resolution of the array improves, but the images are also smaller. Depending how optically thin the model is (how large and negative the spectral index is), the image can be extremely small as is the case for MBQ. For a spectral index similar to that observed by \citet{marronephd}, the image will still be large at $.87$mm, and the shadow should be observable from the visibility. 

To demonstrate this explicitly, visibility amplitude as a function of baseline length is plotted in Fig. \ref{1dchile} for the best fit models, considering baselines made up of combinations of telescopes in California (CARMA), Arizona (SMTO), Mexico (LMT) and Chile (APEX/ASTE/ALMA). The shadow is detectable for the best fit MBD and MBQ models on the Chile--Mexico and Chile-Arizona baselines. The differences in the predictions from the models in Fig. \ref{1dchile} are almost exclusively caused by their different best fit sky orientations.

Since Sgr A* varies significantly on timescales short compared to its mutual visibility between current and future VLBI telescopes, the profiles shown in Fig. \ref{1dchile} will necessarily include source variability. In addition, the short coherence time of the atmosphere and varying signal/noise between telescopes in the arrays may prevent proper calibration of the visibilities \citep{doeleman2009}. In these cases, the closure phase \citep{rogers1974}, the sum of the phases measured along a triangle of baselines, may be used as a non-imaging observable. It is independent of calibration errors, and probes the source structure in the uv-plane. \citet{doeleman2009} used a RIAF+hotspot model to compute predicted closure phases for future VLBI experiments. We make predictions for the best fit models discussed here, whose time-dependent structure comes directly out of the turbulence driven by the MRI. 

Fig. \ref{cphase} shows the closure phase maps vs. observer time and source sky orientation for the current Arizona/California/Hawaii triangle and a future one consisting of Arizona/Mexico/Chile. Due to the moderate inclination favored by the VLBI and spectral observations, the closure phase is relatively robust over time, except over a narrow range of  sky orientations. At these positions, one of the baselines in the triangle is sampling the black hole shadow, which causes a sudden rotation of the phase when the real part of the visibility switches sign. The earth's rotation moves the baseline positions, slowly changing the sky orientations at which this effect is visible. For large phase rotations, the visibility amplitude is usually quite small ($\lesssim .1$Jy) and will be difficult to detect. However, it could be observed as a transition in phase over a short timescale, where Sgr A* is detected at one fairly stable closure phase, goes to an undetectable amplitude, and reappears at a different stable closure phase. 

If the sky orientation of Sgr A* is $\sim0-50^\circ$, this shadow signature would be detectable using the Arizona/Mexico/Chile triangle. The best fit images for MBD, MBQ and 90h in Fig. \ref{1dchile} are in this range. Note that the exact value of the phase rotation and how often it occurs in time are model-dependent -- in Fig. \ref{cphase} it is usually $\sim \pi/2$ and occurs at $1-3$ observer times. Regardless of the value of the position angle, the black hole shadow can be detected from closure phase measurements on some triangle of baselines in this way. Closure phases from a larger set of baselines will constrain the structure of the accretion flow in a similar fashion to the current measurements of the visibility amplitude. For simplicity, the closure phases shown here are for almost co-linear baseline trios. Much more complex and time-dependent closure phase signatures are found on nearly triangular baseline arrangements, such as those including Hawaii, Chile and California/Arizona/Mexico.

\begin{figure}
\epsscale{1.1}
\plotone{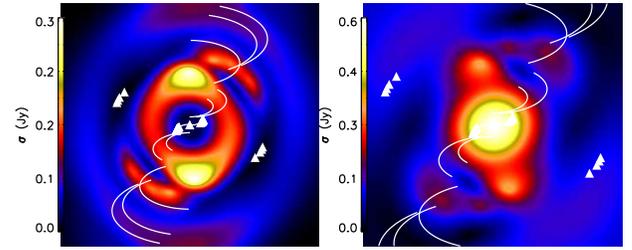}
\caption{\label{sigmav}Standard deviations of the best fit visibility amplitudes at $1.3$mm and $.87$mm. The uv-plane locations of the \citet{doeleman2008} observations are over-plotted as triangles. Future baselines between Chile/Mexico/California/Arizona are over-plotted as solid lines.}
\end{figure}

We calculate the standard deviation of the small sample of best fit visibilities, as done for a large ensemble of RIAF images by \citet{fish2009}. The result from the best fit 3D GRMHD models, shown in Fig. \ref{sigmav}, is in relatively good agreement with their Figure 2. The best baselines for constraining the accretion flow are of similar to slightly shorter length as the existing measurements ($\sim 2000-4000$km), but at roughly orthogonal orientation. The peak values of the variance are larger at $.87$mm. This is expected since we fit to existing measurements at $1.3$mm, while at $.87$mm only the total flux is somewhat constrained by the spectral index between $.4$mm and $1.3$mm from \citet{marronephd}.

\begin{figure}
\epsscale{1.0}
\plotone{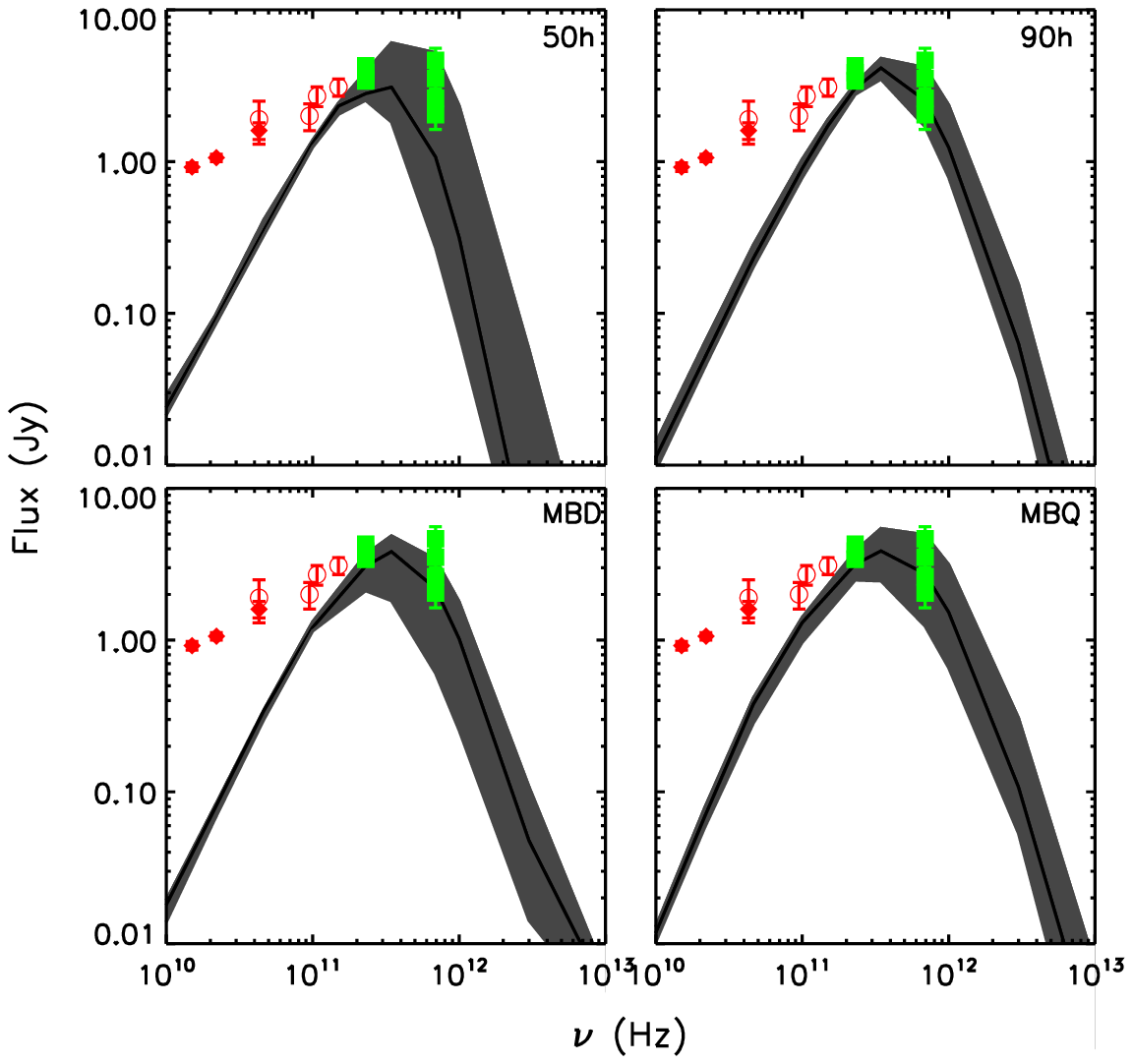}
\caption{\label{spectra}Millimeter spectra for the best fit models. The solid curves are the median values at each frequency, while the dark gray envelope shows the range reached during the simulations. Data points are from \citet{falcke1998} (open circles), \citet{an2005} (filled diamonds) and \citet{marronephd} (filled squares). The models are fit to the \citet{marronephd} data, while at lower frequencies the emission is dominated by emission outside of the simulation domain, and nonthermal emission from electrons either in the accretion flow \citep{yuanquataert2003} or in a short jet \citep{falckemarkoff2000}.}
\end{figure}

The radio spectra from all models are shown in Fig. \ref{spectra}. The shaded gray regions show the extent of the variability at each frequency. Below $\sim 10^{11}$Hz, a substantial portion of the emission is produced outside $r=25$M, and so is not included here. In addition, we do not model emission from nonthermal electrons, and so do not expect to be able to fit the observations outside of the millimeter peak. The variability increases with frequency, due to the decreasing optical depth. The 50h simulation produces a large amount of variability, which may be inconsistent with the range observed from Sgr A*.

\begin{figure}
\epsscale{1.0}
\plotone{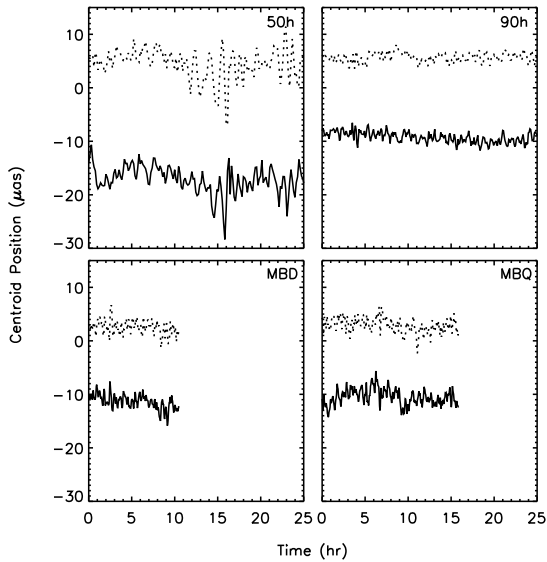}
\caption{\label{centroid}X (solid) and Y (dotted) centroid positions vs. observer time for the best fit models at $345$GHz. The position wander is similar for all frequencies in Fig. \ref{spectra}.}
\end{figure}

In addition to VLBI observations, there are constraints on the time-dependent structure of the accretion flow from astrometry. The centroid wander of Sgr A* at $7$mm was constrained by \citet{reid2008} to be $\lesssim 100\mu$as on timescales of a few hours. Fig. \ref{centroid} shows $x$ and $y$ centroid positions as functions of time for the best fit models at $345$GHz. The position wander for all models considered here is $\lesssim 30\mu$as at millimeter frequencies on all timescales, well within the observed upper limit. A prediction of the simulations is that the millimeter position wander shouldn't exceed that level, even during flares. This is due to the robustness of the time-dependent image structure at all inclinations, and a result of the fact that the variability is produced globally in the inner radii, rather than by small, orbiting inhomogeneities (hotspots).

The instrument GRAVITY at the Very Large Telescope (VLT) will provide $\simeq 15\mu$as precision astrometry of Sgr A* in the infrared (IR). If the IR flaring is due to a relatively global process such as Compton scattering from millimeter seed photons, centroid measurements from thermal synchrotron emission in the IR may be similar to those during the flares. Since the position wander from our models rarely exceeds the detection limit of GRAVITY at any frequency, it should be easy to distinguish such IR flare mechanisms from nonthermal events involving hotspots or expanding blobs, which would lead to larger centroid movements \citep{zamaninasab2010}.

\subsection{Emission region properties}

The millimeter emission region is in the innermost radii of the accretion flow, peaked in the midplane. The emission is dominated by synchrotron emission from thermal electrons \citep{yuanquataert2003}. The saturated magnetic field strengths are typically $\sim .01-.1$ of equipartition, and both the particle density and magnetic pressure are proportional to the accretion rate. These quantities, then, are scaled together to produce the observed total flux from Sgr A*. For these reasons, the quantity that differs most between simulations is the electron temperature. The radial distribution of particle density and the electron temperature in large part determine the image structure and viability of the model.

As expected for accretion disks with aligned angular momentum and black hole spin axes, the effective inner radius of the disk in all simulations moves closer to the black hole with increasing spin.\footnote{For the misaligned case, see \citet{fragiletilt2009}.} However, the density at the event horizon is usually only a factor $\sim 5$ lower than at the peak. There is no sharp inner boundary to the accretion flow.

\subsubsection{Optical Depth Effects}

For best fit parameter combinations, the $1.3$mm photosphere is quite compact and the flow is mostly optically thin, especially at low inclination. In the 0h and 50h simulations, however, as well as for $T_i/T_e=10$ in all models, the photosphere at $1.3$mm extends well outside of the region of peak emissivity, at $r \sim 5$M. In general, the photosphere gets smaller at higher electron temperatures, since those models require lower accretion rates to match the observed flux from Sgr A*. Optical depth effects are also more important at high inclination. This is because Doppler beaming increases the absorption along rays between the observer and the approaching gas.

Single temperature conservative models (MBD, MBQ) have no noticeable photosphere. This leads to a small image size, and low accretion rates. Since the \citet{marronephd} observations find that the turnover in the synchrotron spectrum occurs between $1.3$mm and $.4$mm, they are inconsistent with models that are completely optically thin at $1.3$mm. The VLBI observations also disfavor such models due to their small image sizes. For $T_i/T_e=10$, we have the opposite situation. VLBI rules out these models due to their large photospheres and hence image sizes, while they are generally also optically thick enough that the synchrotron turnover occurs shortward of $.4$mm. For all simulations, the best fit $T_i/T_e$ occurs when there is a  photosphere inside of $r \sim 5$M.

\subsubsection{Comparison to RIAF Models}

There are significant differences between the simulated and semi-analytical RIAF accretion disks. In the RIAF model, the disk is assumed to have a constant scale height $H/R=1$, whereas for all simulations considered here $H/R \sim .1-.3$ in the regions of peak millimeter synchrotron emissivity, as measured by fitting a Gaussian to the particle density at the median radius of peak synchrotron emissivity. The Gaussian fits describe the vertical particle density distribution, and their parameters are stable over time. The emissivity has the same vertical profile.

Thinner disks have higher particle density, stronger magnetic fields and smaller temperatures. These effects are largely mitigated in the Sgr A* models, since the electron temperature scaling is a free parameter, and the magnetic pressure and density scale with accretion rate. The parameter $\beta = P_g/P_m$ is independent of accretion rate and electron temperature. In regions of peak synchrotron emissivity, it has a median value around $\sim 3-10$ for the high spin simulations 90h, MBD and MBQ. The lower spin, non-conservative simulations 0h and 50h have median $\beta \sim 10-20$. In RIAF models, $\beta$ is assumed to be a constant, usually around $\sim 10$. The trend of increasing $\beta$ with decreasing spin is also seen in the axisymmetric HARM simulations. Although $\beta$ varies spatially in the simulations, the emission region is small enough that these variations aren't larger than an order of magnitude, so that constant $\beta$ is likely a good approximation.

We have also fit radial profiles of density, temperature and magnetic field strength to compare the simulations to the RIAF models. In general, simple power laws $r^{-\alpha}$ describe the radial distribution of temperature and magnetic field strength within the millimeter emission region. The power law indices for the temperature distribution are around $\alpha \sim 0.7$, which is quite close to the RIAF value of 0.84. The conservative MBD and MBQ simulations have magnetic field indices of $\alpha \sim 1.1-1.2$, while the 0h, 50h and 90h simulations range from $\alpha \sim 0.3-0.7$. The RIAF value is $1.05$. In the 50h, 90h and 2D HARM simulations, the magnetic field power law index increases with spin.

Other than the disk thickness, the main discrepancy between RIAF models and GRMHD simulations is in the radial particle density profile. The simulations find flat or decreasing density profiles towards the horizon from a point at $\sim 2 r_{ms}$, where $r_{ms}$ is the marginally stable orbit. In RIAF models, the density increases up until the horizon as a fixed power law. This is especially inaccurate at low spin, where the 0h, 50h, and HARM simulations find peaks in the radial density profile at $r \sim 10-15$M, depending on the value of the spin.

Finally, running the axisymmetric HARM simulations also allows a direct comparison to the spectral index values found in \citet{moscibrodzka2009} for different ray tracing codes and interpolation schemes. Values of the spectral index, time-averaged as discussed in that work, are in good agreement with their results in most cases. This is even true when we use the angle-averaged emissivity from \citet{maha}. The emissivity from Leung et al (2010; in preparation) takes into account the angle $\theta$ between the ray and the magnetic field. The largest differences between the emissivities are at extreme inclinations, where a factor of $\sin{\theta}$ in $\nu_c$ causes the spectrum to peak at larger (smaller) frequencies at low (high) inclinations. Individual images and spectra can significantly differ between emissivities. However, all parameter estimates, probability distributions, light curves and best fit models presented here are robust to the choice of emissivity.

\section{Discussion}
\label{discussion}

The submillimeter bump in Sgr A* provides a unique laboratory for connecting observations with  theoretical models of black hole accretion flows. We have performed radiative transfer as a post-processor on the output from existing three dimensional GRMHD simulations and fit them to VLBI and spectral index observations of the Galactic center black hole. The simulations provide excellent fits to the existing millimeter observations, and allow us to estimate the roughly constant temperature of the millimeter emission region and the inclination and sky orientation angles of the black hole. These are found to be $T_e=(5.4 \pm 3.0) \times 10^{10}$ K, $i={50^\circ}^{+35{}^\circ}_{-15{}^\circ}$ and $\xi={-23^\circ}^{+97{}^\circ}_{-22{}^\circ}$ with $90\%$ confidence. The estimates are in excellent agreement with those found by  \citet{broderick2009} for RIAF models, despite significant differences between the dynamical models used. 

The face-on models, which provided excellent fits in the 90h simulation considered in \citet{dexter2009}, are ruled out to $3\sigma$ due to the spectral index constraint in 90h, and the paucity of good VLBI and spectral fits at low inclinations in the other simulations. For this reason, it may not be possible to detect the shadow of the black hole from visibility measurements using the telescopes from \citet{doeleman2008}. The best fit images from our modeling are crescents, with the shadow unobscured in all viable models. The shadow may be detectable with measurements on a baseline between Chile and Mexico, California or Arizona either by directly sampling the visibility profile, or through observations of the closure phase. These same baselines are also the most promising for further constraining black hole parameters, and testing accretion flow models. Finally, our accretion rate constraint of $\dot{M}=5^{+15}_{-2}\times10^{-9} M_\odot \mathrm{yr}^{-1}$ is consistent with but independent of estimates from the observed linear polarization and Faraday rotation measures \citep{agol2000,quataert2000,marrone2007}.

Millimeter light curves from all best fit models show flaring events with $\sim 50\%$ amplitudes and few hour durations at a frequency of $1-2$ per day, consistent with the millimeter flaring behavior of Sgr A*. These flares are caused by rises in magnetic field strength near the midplane of the inner radii of the accretion flow, due to magnetic turbulence driven by the MRI. They do not appear to be linked to heating from magnetic reconnection, and are accurately reproduced by a simple model assuming an isothermal emission region with $\nu/\nu_c \sim 20$. The variability in the light curves for all simulations at $.4$mm is strongly correlated with that in the accretion rate onto the black hole. However, there is no clear evidence for a characteristic lag between the two. 

Due to the uncertainties in their emission mechanisms and electron distribution functions, we do not model the radio, IR or X-ray portions of the spectrum, and so are unable to test for the presence of correlated multiwavelength flares. \citet{maitra2009} showed that the observed time lags between the radio and millimeter flares may be reproduced by a jet model, where perturbations in the accretion rate near the black hole expand as they flow outward. Since the millimeter variability is strongly correlated with the accretion rate, that mechanism could explain the correlations between radio and millimeter flares even if the millimeter emission is produced in an accretion disk rather than an outflow.

The sample of simulations used does not span the parameter space of black hole spin, initial torus geometry and initial magnetic field configuration. This incompleteness may affect our parameter estimates. However, the universality of the images, light curves, probability distributions and emission region characteristics from all viable simulations considered here suggests that the changes will probably be minor. The disk thickness is especially important, since the emission region scale height mimics that of the disk. A thicker disk will also have larger MHD temperatures, and thus larger values of $T_i/T_e$ will be necessary to fit the observations. If the disk thickness can be constrained by VLBI observations, a constraint on $T_i/T_e$ follows. This parameter would constrain the strength of ion-electron coupling in collisionless plasmas. Simulations with misaligned angular momentum and black hole spin axes \citep{fragile2007,fragile2008,fragileetal2009} are also of interest, since they produce standing shocks which lead to considerably different image morphologies. Parameters such as the inclination angle or spin may be degenerate with the tilt angle. Fits from these models will be considered in a subsequent publication.

\citet{broderick2009} also estimated the parameters of Sgr A* from VLBI and spectral constraints by fitting semi-analytic RIAF models. The structure of $p$($i$) and $p$($\xi$) in their Figure 7 (middle and right panels) are in excellent agreement with ours in Figs. \ref{pvi} \& \ref{pvx}. Further, we find that future observations on baselines of similar length but orthogonal orientation to the existing Hawaii-Arizona baseline have the best chance of discriminating between our best fit models. This conclusion is in agreement with \citet{fish2009}, who point out that ideal baselines for such measurements would be between Chile and one or more of California, Arizona or Mexico. A simultaneous measurement of the 345 or 690 GHz flux during future 230 GHz VLBI observations would provide a significant additional constraint.

A grid of time-averaged images and spectra over black hole spin, inclination and $T_i/T_e$ from axisymmetric, total energy conserving GRMHD simulations were compared to millimeter observations by \citet{moscibrodzka2009}. Their finding that $T_i/T_e=1$ models are inconsistent with observations agrees with our results here for MBD and MBQ. As we have shown, for simulations with lower MHD temperatures, smaller values of $T_i/T_e$ will be necessary to reach the required electron temperature in the millimeter emission region. This is why the non-conservative simulations considered here (0h, 50, 90h) are all best fit by small values of $T_i/T_e$. It also suggests that such models are only appropriate for modeling Sgr A* when they produce large enough MHD temperatures that the electron temperature lies within the range estimated here. At low spin, this seems to require either a hotter initial condition or a total energy conserving simulation.

In all but the ``best-bet" model from \citet{moscibrodzka2009}, $T_i/T_e=10$ is required for consistency with multiwavelength spectral observations. Millimeter spectral index results from our own HARM simulations modeled in the same fashion are for the most part in good agreement with theirs, indicating that the results of these studies are probably robust to the ray tracing code and interpolation scheme employed. However, $T_i/T_e=10$ models for the various HARM models, MBD and MBQ are all too large to fit the VLBI data due to their extended photospheres. In addition, we find good fits at $T_i/T_e=3$ for MBD and MBQ at moderate to edge-on inclinations. We have not computed IR or X-ray emission, and therefore do not fit to the quiescent X-ray luminosity from \citet{baganoff2003}, or to the upper limits to the IR emission. These limits are violated in \citet{moscibrodzka2009} for nearly edge-on inclinations and at high spin. Including them could disfavor high inclinations, but is unlikely to change our best fit models.

Additional constraints may come from the degree and orientation of linear polarization first detected by \citet{aitken2000}. \citet{huang2009} modeled the polarization from a pseudo-Newtonian model. It is important to use relativistic simulations for such studies, since the degree of polarization is strongest close to the black hole. Future VLBI observations may make use of the closure phase between triangles of baselines to probe changing structures in the accretion flow \citep{doeleman2009}. \citet{fish2009b} extended this work to include polarization. Both of these papers used a RIAF+hotspot model. We have computed the first closure phase signatures from more physically realistic GRMHD simulations, and find that the signature of the black hole shadow is present in the predicted closure phases. Detailed comparison of observed and predicted closure phases will be able to constrain the accretion flow in the same fashion as direct sampling of the visibility amplitude.

\begin{acknowledgements}
J.D. thanks Shep Doeleman for the VLBI data. This work was partially supported by NSF grant AST 0807385, NASA grant 05-ATP05-96, NASA Earth \& Space Science Fellowship NNX08AX59H (JD), and NASA Chandra Fellowship PF7-80048 (JCM).
\end{acknowledgements}

\end{document}